\documentclass[showpacs,preprint]{revtex4}
\usepackage{graphicx}
\usepackage{subfigure}
\usepackage{amssymb}
\usepackage{amsmath}
\usepackage{bm}
\usepackage{latexsym}
\usepackage{hyperref}

\begin {document}

\title{Controllable dynamics of two separate qubits in Bell states}

\author{Jun Jing$^{1}$\footnote{Email address: jingjun@shu.edu.cn},
Zhi-guo L\"{u}$^2$\footnote{Email address: zglv@sjtu.edu.cn},
Guo-hong Yang$^1$}

\affiliation{$^1$Department of Physics, Shanghai University,
Shanghai 200444, China\\ $^2$Department of Physics, Shanghai
Jiaotong University, Shanghai 200240, China}

\date{\today}

\begin{abstract}
The dynamics of entanglement and fidelity for a subsystem of two
separate spin-1/2 qubits prepared in Bell states is investigated.
One of the subsystem qubit labelled $A$ is under the influence of
a Heisenberg XY spin-bath, while another one labelled $B$ is
uncoupled with that. We discuss two cases: (i) the number of bath
spins $N\rightarrow\infty$; (ii) $N$ is finite: $N=40$. In both
cases, the bath is initially prepared in a thermal equilibrium
state. It is shown that the time dependence of the concurrence and
the fidelity of the two subsystem qubits can be controlled by
tuning the parameters of the spin-bath, such as the anisotropic
parameter, the temperature and the coupling strength with qubit
$A$. It is interesting to find the dynamics of the concurrence is
independent of four different initial Bell states and that of the
fidelity is divided into two groups.

\end{abstract}

\maketitle

\section{Introduction}

Entanglement, which exhibits a very peculiar correlation among the
degrees of freedom of a single particle or the distinct parts of a
composite system, is the most intriguing feature of quantum
composite system and a vital resource for quantum computation and
quantum communication \cite{Nielsen, Bennett, Horodecki, Werner,
Greenberger, Dur}. In the field of quantum information theory, it
is a fundamental issue to create, quantify, control and manipulate
the entangled quantum bits, which are often composed of spin-half
atoms in different problems \cite{Nielsen, Loss, Kane, Tanas}.
Particularly, lots of works \cite{Jordan, Benjamin, Barnum} are
devoted to steering two initially entangled qubits through an
auxiliary particle or field (for instance, another spin qubit or a
bosonic mode), which interacts with only one of them. It is a very
exciting motivation, yet those approximated models neglect the
actual complex environment of the quantum qubits. And it remains
an important open question how the entanglement degree responds to
the influence of environmental noise \cite{Yu4}. Practically, the
spin qubits are indeed open quantum subsystems and exposed to the
influence of their environments \cite{Breuer3, weiss, Breuer,
Yuan}. In most conditions, the coupling between the subsystem and
environment will degrade the entanglement degree between the
subsystem qubits. In some other conditions, however, a specially
structured and well designed bath can be conceived as a protection
device to suppress the negative influence from itself or other
noise sources \cite{TWmodel, Milburn, Jing2}. \\

For the spin subsystem, there are two important modes of bath: (i)
boson-bath, e.g., the Caldeira-Leggett model \cite{Leggett}; (ii)
spin-bath, e.g., the model used in Ref. \cite{Prokoev}. Here we
discuss the latter one. It is well known that the localized spins
in solid state nano-devices, the most promising candidates for
qubits due to their easy scalability and controllability
\cite{Loss, Burkard}, are mainly subject to the influence from the
nuclear spins, which constitute a type of spins-1/2 environment.
It is an almost intractable computation task to deal with such a
spin-spin-bath model for its giant number of degrees of freedom.
Therefore, physicists resort to some approximations or
simplifications, such as the Markovian \cite{Gardiner} schemes and
the non-Markovian ones \cite{Shresta}, which have been developed
in the past two decades. Based on these schemes, plenty of
analytical and numerical methods were exploited to study the
reduced dynamics of subsystem consisted of spins-1/2 by tracing
out the degrees of freedom of the spin-bath. Some recent works
focused on the center spins in a network configuration, in which
the form of bath is specially structured, such as a thermal bath
\cite{Paganelli, Paganelli2}, a bath via Heisenberg XX couplings
\cite{Hutton} and a thermal spin bath via Heisenberg XY couplings
\cite{Yuan, Jing3}. \\

In this paper, an open two-spin-qubit subsystem (two qubits
labelled $A$ and $B$ respectively) is explored as a target quantum
information device with a spin bath of a star-like configuration.
The model is something like the one considered in Ref. \cite{Yuan,
Hamdouni}. But there are significant differences between them. It
is supposed that at the beginning, the subsystem is prepared as
one of the Bell states (Einstein-Rosen-Podolsky pairs) \cite{EPR,
Bell}:
\begin{eqnarray*}
|e_1\rangle &=& 1/\sqrt{2}(|11\rangle+|00\rangle) \\
|e_2\rangle &=& 1/\sqrt{2}(|10\rangle+|01\rangle) \\
|e_3\rangle &=& 1/\sqrt{2}(|11\rangle-|00\rangle) \\
|e_4\rangle &=& 1/\sqrt{2}(|10\rangle-|01\rangle)
\end{eqnarray*}
Then Qubit $B$ is moved far away or isolated so that not only the
coupling between Qubit $B$ and $A$, but also the interaction of
$B$ with the spin-bath could be neglected. The bath is regarded as
an adjustable auxiliary device to control the time evolution of
the two qubits. And the evolution is represented by the
concurrence and fidelity dynamics as functions of various
parameters associated with the bath. The reduced dynamics of the
subsystem is obtained by a numerical scheme combined by the
Holstein-Primakoff transformation and the Laguerre polynomial
expansion algorithm. We consider two conditions, in which the
number of spins in the bath is infinite and finite. The rest of
this paper is organized as following. In Sec. \ref{MandM} we first
give the model Hamiltonian and its analytical derivation; and then
we introduce the numerical calculation about the evolution of the
reduced matrix for the subsystem. Detailed results and discussions
are in Sec. \ref{discussion}. And conclusion is given in Sec.
\ref{conclusion}.

\section{Model and Method}\label{MandM}

\subsection{Hamiltonian}\label{Hamiltonian}

The subsystem is consisted of two entangled spin-1/2 atoms
labelled $A$ and $B$ respectively, between which there is no
coupling. Qubit $A$ interacts with a spin-1/2 bath via a
Heisenberg XY interaction while $B$ does not. The total system
Hamiltonian, similar to those considered in Refs. \cite{Yuan,
Breuer, Hutton, Breuer2}, is divided into $H_s$, $H_b$ and
$H_{sb}$. They represent the subsystem, the bath and the
interaction \cite{Breuer, Yuan, Canosa} part between the former
two terms respectively:
\begin{eqnarray}\label{HS}
H_s&=&\mu_0(\sigma_{A}^z+\sigma_{B}^z),\\ \label{HSB}
H_{sb}&=&\frac{g_0}{2\sqrt{N}}\sum_{i=1}^N\left[(1+\gamma)
\sigma_{A}^x\sigma_i^x+(1-\gamma)\sigma_{A}^y\sigma_i^y\right],\\
\label{HB} H_b&=&\frac{g}{2N}\sum_{i\neq
j}^N\left[(1+\gamma)\sigma_i^x\sigma_j^x+
(1-\gamma)\sigma_i^y\sigma_j^y\right].
\end{eqnarray}
where $\mu_0$ is half of the energy bias for the two-level atom
$A$ or $B$; $g_0$ is the coupling strength between the subsystem
spin $A$ and the bath spins; $g$ represents the mutual
interactions among the bath spins. $\gamma$ ($0\leq\gamma\leq1$)
is the anisotropic parameter. When $\gamma=0$, the XY interaction
is reduced to an XX one \cite{Yuan}. The x, y and z components of
the matrix $\sigma$ are the Pauli matrices.
\begin{equation}
\sigma^x=\left(\begin{array}{cc}
      0 & 1 \\
      1 & 0
    \end{array}\right), \quad
\sigma^y=\left(\begin{array}{cc}
      0 & -i\\
      i & 0
    \end{array}\right), \quad
\sigma^z=\left(\begin{array}{cc}
      1 & 0\\
      0 & -1
    \end{array}\right).
\end{equation}
The indices $i$ in the summation run from $1$ to $N$ and $N$ is
the number of the bath spins.\\

Adopting $\sigma^\pm=1/2(\sigma^x\pm i\sigma^y)$, we can rewrite
the Hamiltonians (\ref{HSB}) and (\ref{HB}) by:
\begin{equation}\label{HSBp1}
H_{sb}=\frac{g_0}{\sqrt{N}}\biggl[\sum_{i=1}^N\sigma_i^+(\gamma\sigma_{A}^++
\sigma_{A}^-)+\sum_{i=1}^N\sigma_i^-(\sigma_{A}^++\gamma\sigma_{A}^-)\biggr],
\end{equation}
\begin{equation}\label{HBp1}
H_b=\frac{g}{N}\sum_{i\neq
j}^N\left[\gamma(\sigma_i^+\sigma_j^++\sigma_i^-\sigma_j^-)+
(\sigma_i^+\sigma_j^-+\sigma_i^-\sigma_j^+)\right].
\end{equation}

Substituting the collective angular momentum operators
$J_{\pm}=\sum_{i=1}^N\sigma_{i}^\pm$ \cite{Ficek} into Eqs.
(\ref{HSBp1}) and (\ref{HBp1}), we get
\begin{eqnarray}\label{H_SB1}
H_{sb}&=&\frac{g_0}{\sqrt{N}}\left[J_+(\gamma\sigma_{A}^++
\sigma_{A}^-)+J_-(\sigma_{A}^++\gamma\sigma_{A}^-)\right],\\
\label{H_B1}
H_b&=&\frac{g}{N}\left[\gamma\left(J_+J_++J_-J_-\right)+
\left(J_+J_-+J_-J_+-N\right)\right].
\end{eqnarray}

By the Holstein-Primakoff transformation \cite{Holstein},
\begin{equation}\label{J}
J_+=b^+(\sqrt{N-b^+b}), \hspace*{2mm} J_-=(\sqrt{N-b^+b})b,
\end{equation}
with $[b,b^+]=1$ and the first order approximation of $1/N$,
$$\sqrt{\frac{N-b^+b}{N}}=1-\frac{b^+b}{2N}=1-\frac{\hat{n}}{2N},$$
the Hamiltonians Eq. (\ref{H_SB1}) and Eq. (\ref{H_B1}) can be
written as:
\begin{equation}\label{H_SB}
H_{sb}=g_0\left[(b^+-\frac{b^+\hat{n}}{2N})(\gamma\sigma_{A}^++
\sigma_{A}^-)+(b-\frac{\hat{n}b}{2N})(\sigma_{A}^++\gamma\sigma_{A}^-)\right],
\end{equation}
\begin{equation}\label{H_B}
\begin{split}
H_b=&g\biggr\{\gamma\left[(b^+-\frac{b^+\hat{n}}{2N})(b^+-\frac{b^+\hat{n}}{2N})+
(b-\frac{\hat{n}b}{2N})(b-\frac{\hat{n}b}{2N})\right]
\\   &+
(b^+-\frac{b^+\hat{n}}{2N})(b-\frac{\hat{n}b}{2N})+
(b-\frac{\hat{n}b}{2N})(b^+-\frac{b^+\hat{n}}{2N})-1\biggr\}
\\=&g\biggr\{\gamma\left[(b^+)^2(1-\frac{\hat{n}+1}{2N})(1-\frac{\hat{n}}{2N})
+b^2(1-\frac{\hat{n}-2}{2N})(1-\frac{\hat{n}-1}{2N})\right]
\\
&+2\hat{n}+\frac{\hat{n}(2\hat{n}^2-8N\hat{n}-\hat{n}+1)}{4N^2}\biggr\}.
\end{split}
\end{equation}

Utilizing the collective environment pseudospin $J$ and the
Holstein-Primakoff transformation, one could reduce a
high-symmetric spin bath, such as the one we considered, into a
single-mode bosonic bath field \cite{Breuer, Yuan}. The
transformed Hamiltonian is just like a spin-boson model in the
field of cavity quantum electrodynamics (CQED). And the effect of
the single-mode bath on the dynamics of the two subsystem qubits
is interesting although the bath only directly interacts with one
of them. The model might be helpful to understand the magic
essence of quantum entanglement and practical in manipulating the
quantum communication.

\subsection{Calculation method}\label{Cmethod}

The whole state of the total system is assumed to be separable
before $t=0$, i.e.
$\rho(0)=|\psi(0)\rangle\langle\psi(0)|\otimes\rho_b$. The
subsystem $|\psi(0)\rangle$ is prepared as one of the four Bell
states, $|e_{i}\rangle, i=1,2,3,4$. The bath is in a thermal
equilibrium state, $\rho_b(0)=e^{-H_b/k_BT}/Z$, where $Z={\rm
Tr}\left(e^{-H_b/k_BT}\right)$ is the partition function. The
Boltzmann constant $k_B$ is set to be $1$ for the sake of
simplicity in later calculation. To derive the density matrix
$\rho(t)$ of the whole system,
\begin{equation}\label{rhot}
\rho(t)=\exp(-iHt)\rho(0)\exp(iHt),
\end{equation}
we need to consider two factors. \\

(i) To express the thermal bath, we use the method suggested by
Tessieri and Wilkie \cite{TWmodel, Jing2, Jing3}:
\begin{eqnarray}\label{rho_B}
\rho_b(0)&=&\sum_{m=1}^{N}|\phi_m\rangle\omega_m\langle\phi_m|,\\
\omega_m&=&\frac{e^{-E_m/T}}{Z},\\ \label{weight}
Z&=&\sum_{m=1}^{N}e^{-E_m/T}.
\end{eqnarray}
where $|\phi_m\rangle$, $m=1, 2, 3, \cdots, N$, are the
eigenstates of the environment Hamiltonian $H_b$, and $E_m$ the
corresponding eigen energies in increasing order. On the condition
of thermodynamics limit, i.e. $N\rightarrow\infty$, Eq.
(\ref{H_SB}) and Eq. (\ref{H_B}) are simplified as:
\begin{eqnarray}\label{Hsbp2}
H_{sb}&=&g_0\left[b^+(\gamma\sigma_{A}^++
\sigma_{A}^-+\gamma\sigma_{B}^++\sigma_{B}^-)+b(\sigma_{A}^++\gamma\sigma_{A}^-+
\sigma_{B}^++\gamma\sigma_{B}^-)\right],\\ \label{Hbp2}
H_b&=&g[\gamma({b^+}^2+b^2)+ 2b^+b].
\end{eqnarray}
Then $N$ in Eq. (\ref{rho_B}) and Eq. (\ref{weight}) should be
replaced with a cutoff $M$ linking to a certain high energy level.
By the above expansion, the initial state can be represented by:
\begin{equation}\label{rho0}
\rho(0)=\sum\omega_m|\Psi_m(0)\rangle\langle\Psi_m(0)|, \quad
|\Psi_m(0)\rangle =|\psi(0)\rangle|\phi_m\rangle.
\end{equation}

(ii) For the evaluation of the evolution operator
$U(t)=\exp(iHt)$, we apply the Laguerre polynomial expansion
scheme, which is proposed by us \cite{Jing2, Jing3, Jing}, into
the computation.
\begin{equation}
U(t)=\left(\frac{1}{1+it}\right)^{\alpha+1}
\sum^{\infty}_{k=0}\left(\frac{it}{1+it}\right)^kL^{\alpha}_k(H).
\end{equation}
$L^{\alpha}_k(H)$ is one type of Laguerre polynomials
\cite{Arfken} as a function of $H$, where $\alpha$
($-1<\alpha<\infty$) distinguishes different types of the Laguerre
polynomials and $k$ is the order of them. In real calculations the
expansion has to be cut at some value of $k_{\text{max}}$, which
was optimized to be $20$ in this study (We have to test out a
$k_{\text{max}}$ for the compromise of the numerical stability in
the recurrence of the Laguerre polynomial and the speed of
calculation). With the largest order of the expansion fixed, the
time step $t$ is restricted to some value in order to get accurate
results of the evolution operator. At every time step, the
accuracy of the results will be confirmed by the test of the
numerical stability --- whether the trace of the density matrix is
$1$ with error less than $10^{-12}$. For longer time, the
evolution can be achieved by more steps. The action of the
Laguerre polynomial of Hamiltonian to the states is calculated by
recurrence relations of the Laguerre polynomial. The scheme is of
an efficient numerical algorithm motivated by Ref.
\cite{Dobrovitski1, Hu} and is pretty well suited to many quantum
problems, open or closed. It could give results in a much shorter
time compared with the traditional methods, such as the well-known
$4$-order Runge-Kutta algorithm, under the same requirement of
numerical accuracy. \\

After some derivations, the density matrix of the whole system
$\rho(t)$ can be determined by Eqs. \ref{rhot} and \ref{rho0}.
Tracing out the degrees of freedom of the environment, we finally
obtain the dynamics of the subsystem qubits:
\begin{equation}
\rho_s(t)={\rm Tr}_b\left(\rho(t)\right).
\end{equation}

\section{Simulation results and discussions}\label{discussion}

With $\rho_s(t)$, we can discuss: (i) the concurrence
\cite{Wootters1, Wootters2}, which is a very good measurement for
the intra-entanglement of two two-level particles and defined as:
\begin{equation}\label{Concurrence}
C=\max\{\lambda_1-\lambda_2-\lambda_3-\lambda_4,~0\},
\end{equation}
where $\lambda_i$, $i=1,2,3,4$, are the square roots of the
eigenvalues of the product matrix
$\rho_s(\sigma^y\otimes\sigma^y)\rho^*_s(\sigma^y\otimes\sigma^y)$
in decreasing order; (ii) the fidelity \cite{Privman}, which is
defined as
\begin{equation}\label{fide}
Fd(t)={\rm Tr}[\rho_{\rm ideal}(t)\rho_s(t)].
\end{equation}
$\rho_{\rm ideal}(t)$ represents the pure state evolution of the
subsystem under $H_s$ only, without interaction with the
environment. The fidelity is a measure for decoherence and depends
on $\rho_{\rm ideal}$. It achieves its maximum value $1$ only if
the time dependent density matrix $\rho_s(t)$ is equal to
$\rho_{\rm ideal}(t)$. \\

The results and discussions about the two quantities are divided
into two subsections, \ref{infinity} and \ref{finite}. Both of
these two cases suggest a type of controlling method to the
entangled qubits, especially the latter one. \\

It is interesting to note that no matter which one of the four Bell
states is chosen as the initial state for the subsystem, there is no
distinction between their dynamics of concurrence. So we need not to
concretely point out the initial states in the following
discussions. But there are differences between the fidelity dynamics
of $|e_{1,3}\rangle$ and $|e_{2,4}\rangle$. It is shown that the
spins in the initial state of the $1$st group, $|e_1\rangle$ and
$e_3\rangle$, are parallel; while in that of the $2$nd group,
$|e_2\rangle$ and $e_4\rangle$, are antiparallel. That is why the
two groups have different fidelity evolution behaviors. In fact, any
other basis obtained from the Bell states by replacing the
coefficients $\pm1/\sqrt{2}$ with $e^{i\theta}/\sqrt{2}$ ($\theta$
being a real number) has the same dynamics \cite{gedik}.

\subsection{Thermodynamic limit ($N\rightarrow\infty$)}\label{infinity}

\begin{figure}[htbp]
\centering \subfigure[$C(t)$]{\label{bellga:C}
\includegraphics[width=3in]{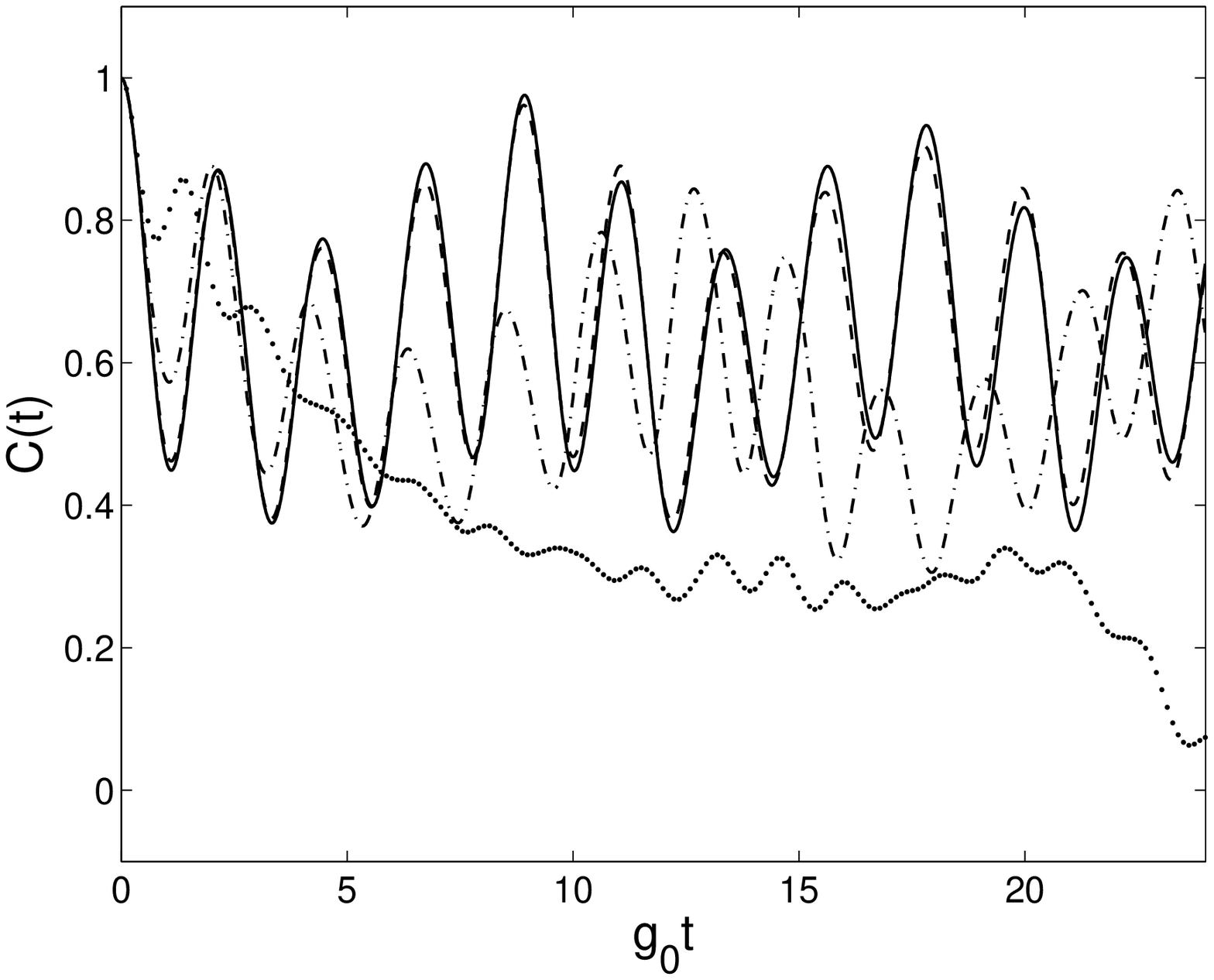}}
\subfigure[$Fd1(t)$]{\label{bellga:Fd0011}
\includegraphics[width=3in]{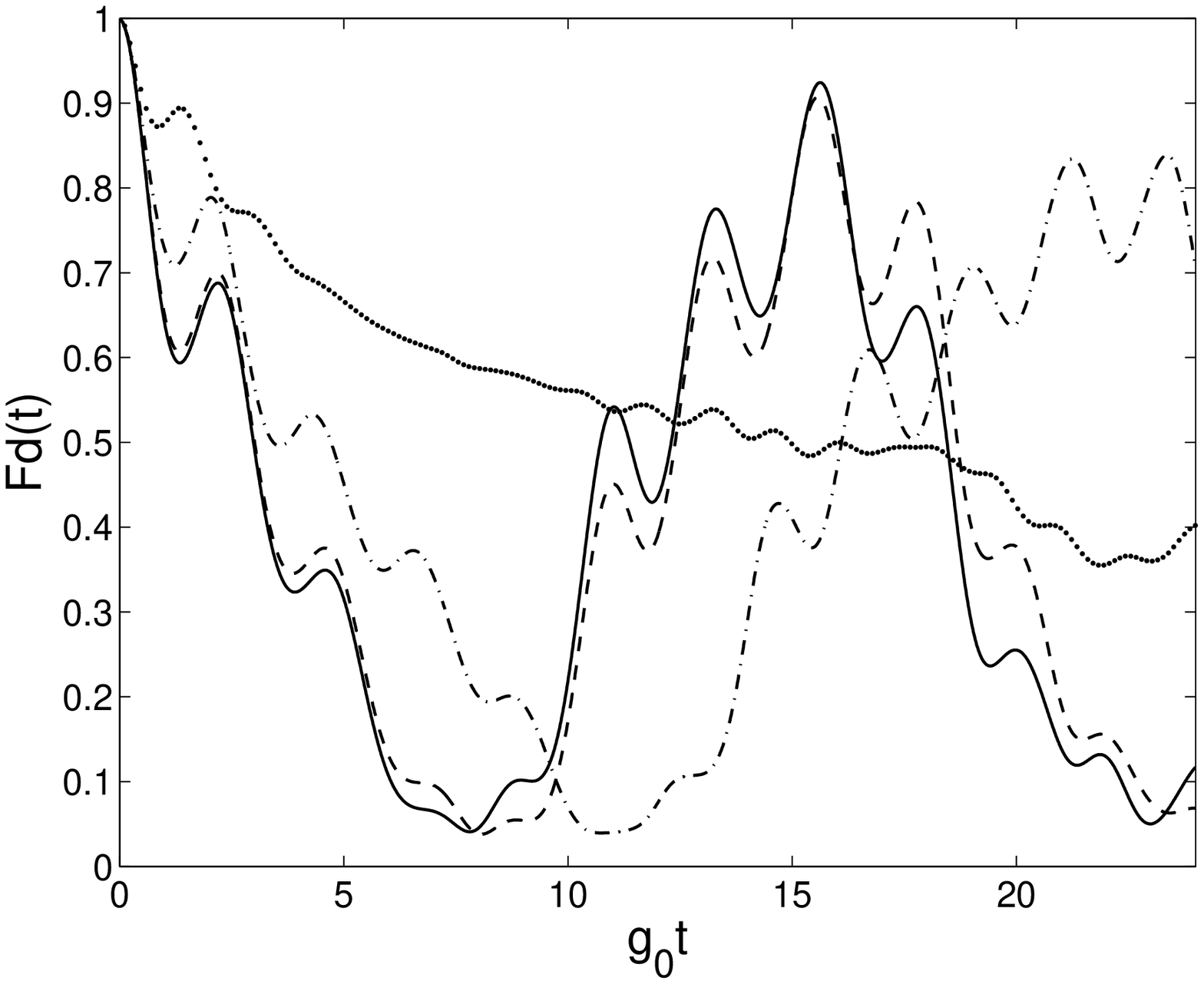}}
\subfigure[$Fd2(t)$]{\label{bellga:Fd1001}
\includegraphics[width=3in]{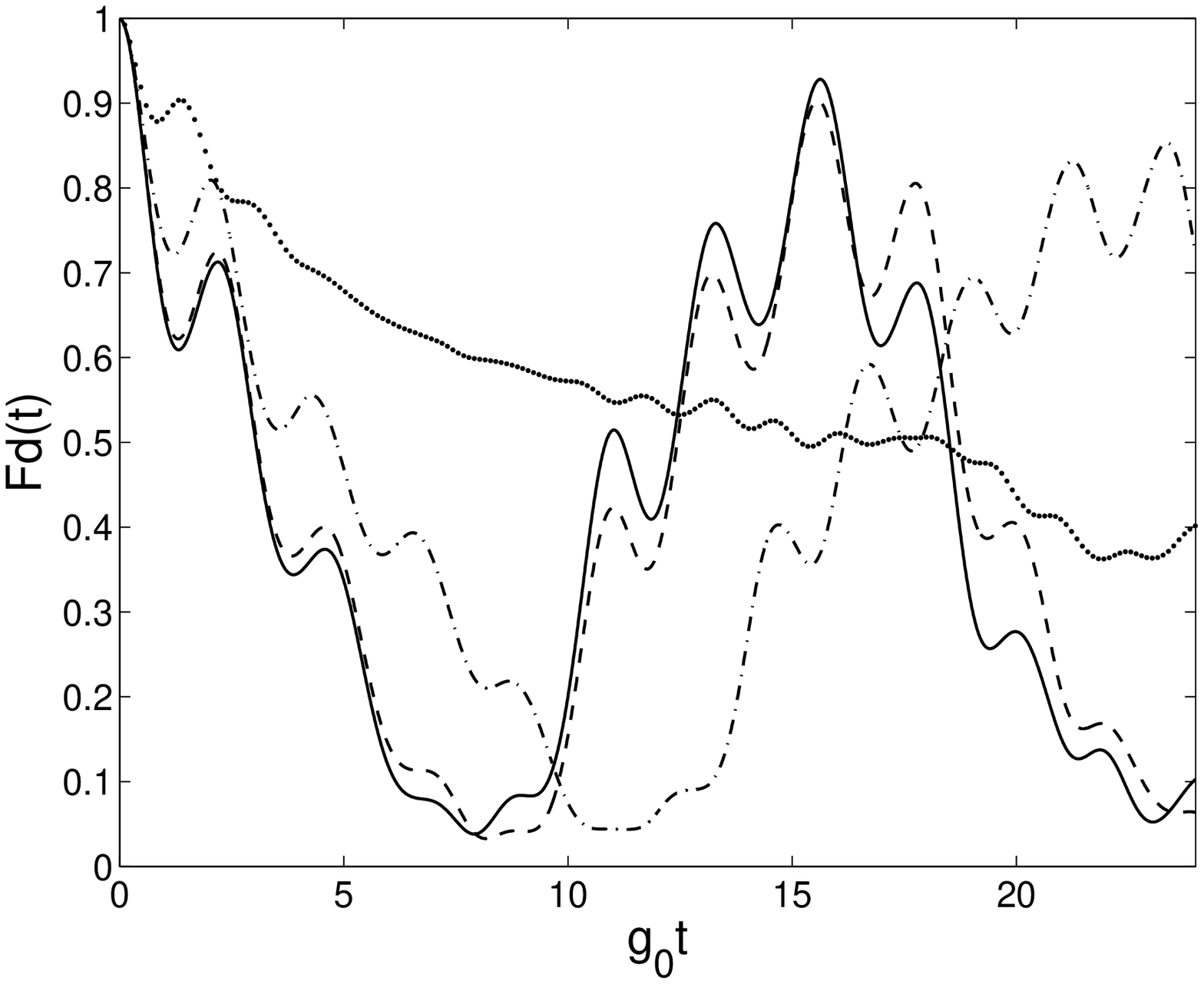}}
\caption{Time evolution of (a) Concurrence from all $4$ Bell
states, (b) Fidelity from $|e_{1,3}\rangle$, (c) Fidelity from
$|e_{2,4}\rangle$ at different anisotropic parameter: $\gamma=0$
(solid curve), $\gamma=0.2$ (dashed curve), $\gamma=0.6$ (dot
dashed curve), $\gamma=1.0$ (dotted curve). Other parameters are
$N\rightarrow\infty$, $\mu_0=2g_0$, $g=g_0$, $T=g_0$.}
\label{bellga}
\end{figure}

\begin{figure}[htbp]
\centering \subfigure[$C(t)$]{\label{bellT:C}
\includegraphics[width=3in]{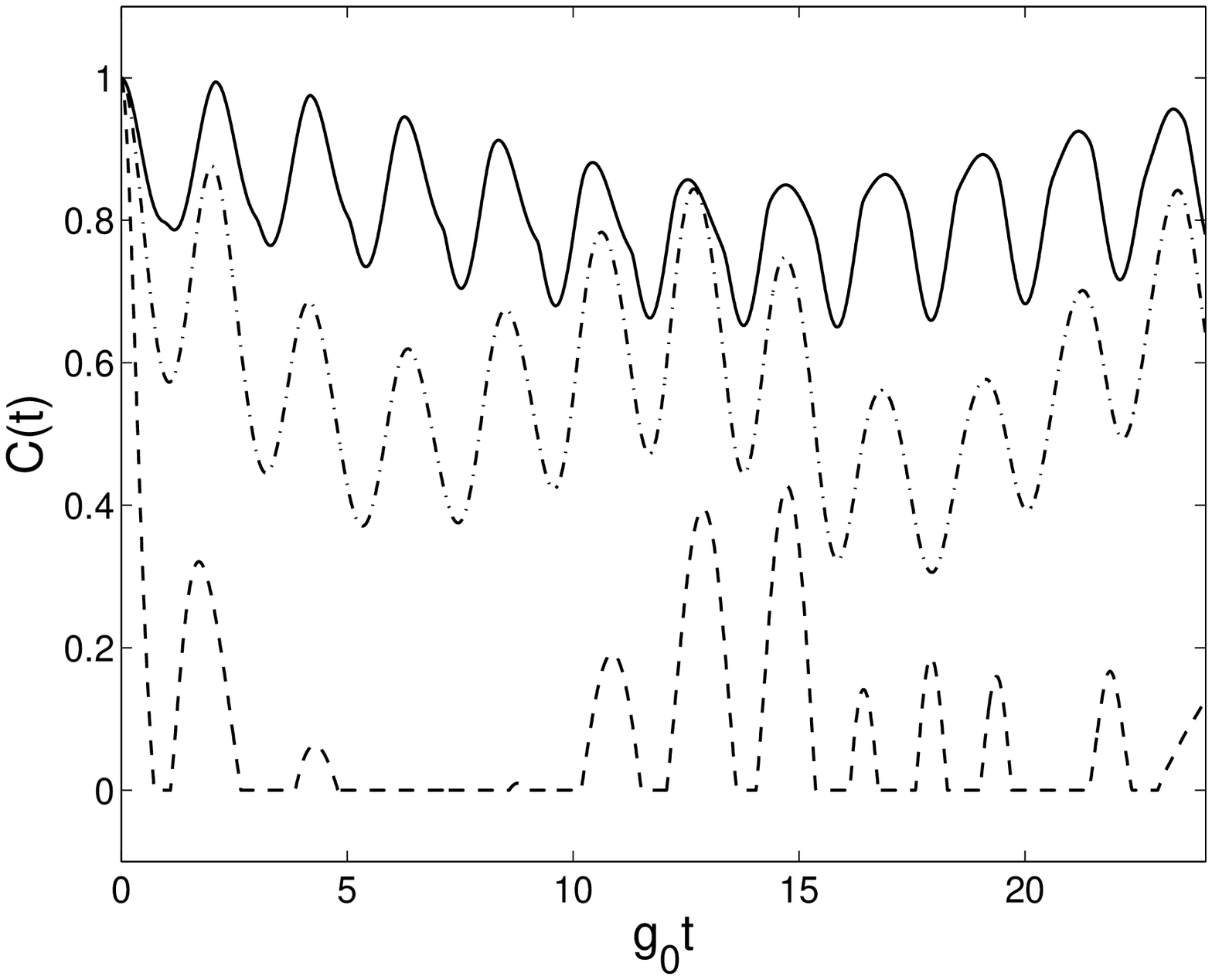}}
\subfigure[$Fd1(t)$]{\label{bellT:Fd0011}
\includegraphics[width=3in]{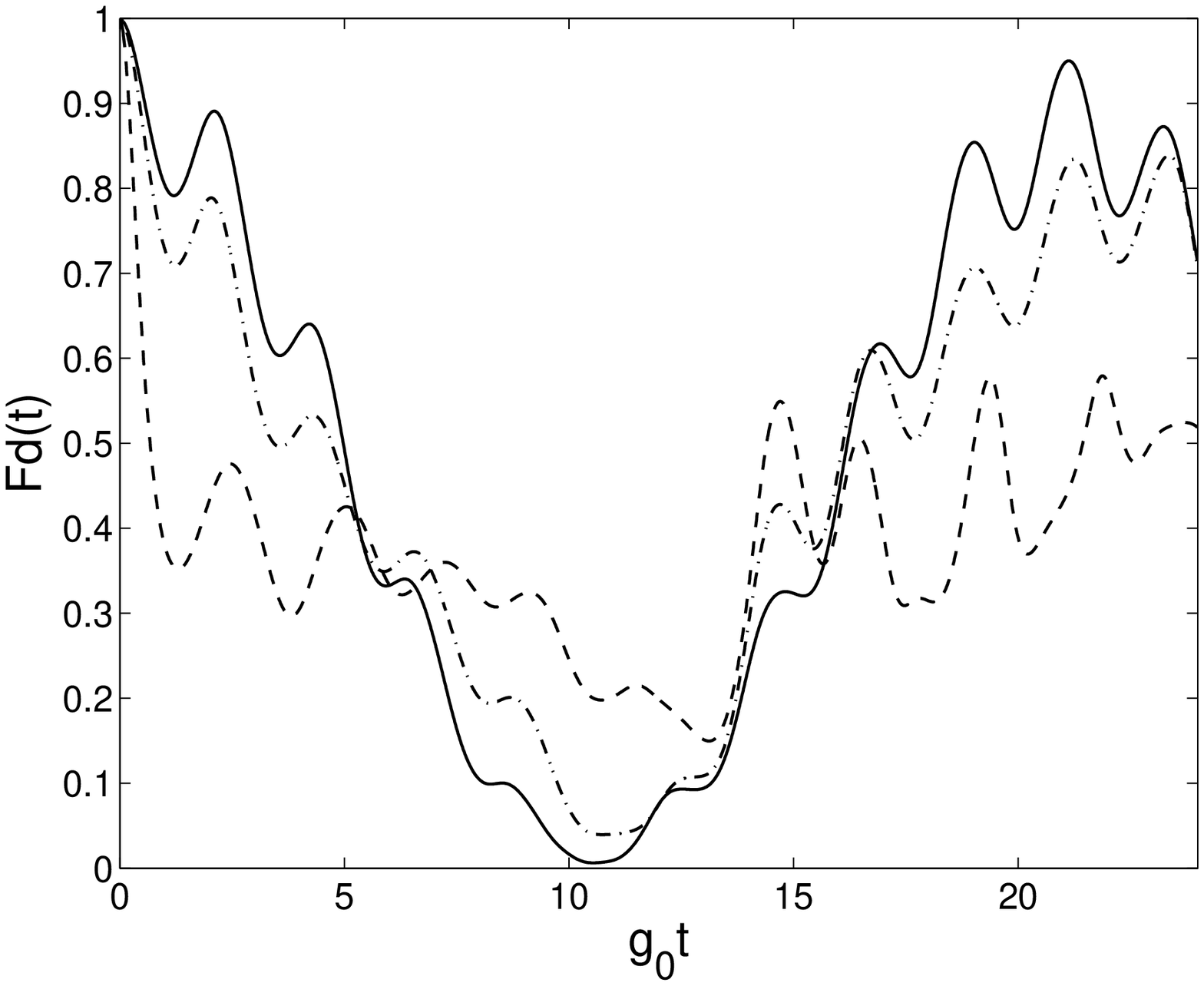}}
\subfigure[$Fd2(t)$]{\label{bellT:Fd1001}
\includegraphics[width=3in]{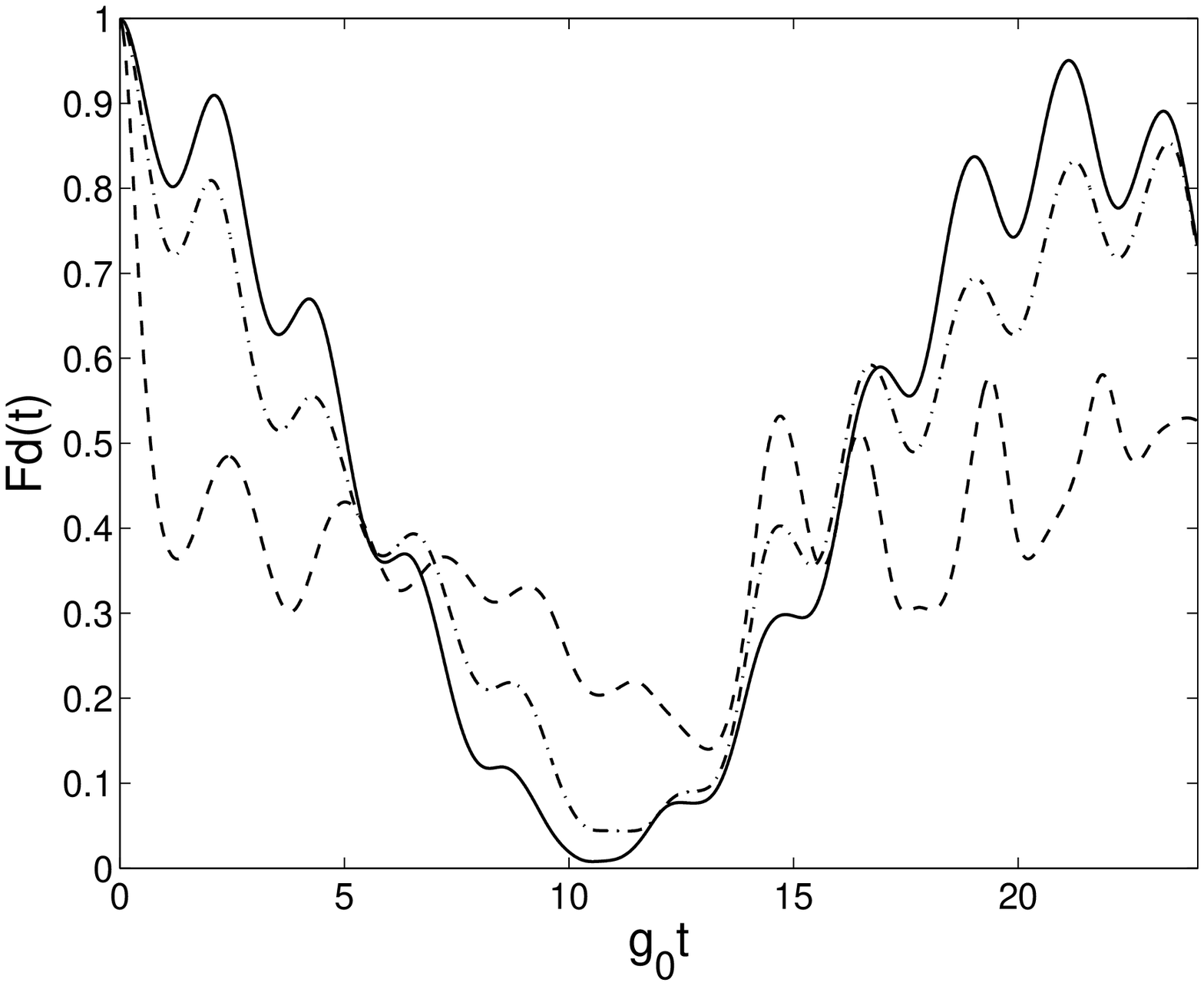}}
\caption{Time evolution for (a) Concurrence from all $4$ Bell
states, (b) Fidelity from $|e_{1,3}\rangle$, (c) Fidelity from
$|e_{2,4}\rangle$ at different values of temperature: $T=0.2g_0$
(solid curve), $T=g_0$ (dot dashed curve), $T=5g_0$ (dashed
curve). Other parameters are $N\rightarrow\infty$, $\mu_0=2g_0$,
$g=g_0$, $\gamma=0.6$.} \label{bellT}
\end{figure}

\begin{figure}[htbp]
\centering \subfigure[$C(t)$]{\label{bellg:C}
\includegraphics[width=3in]{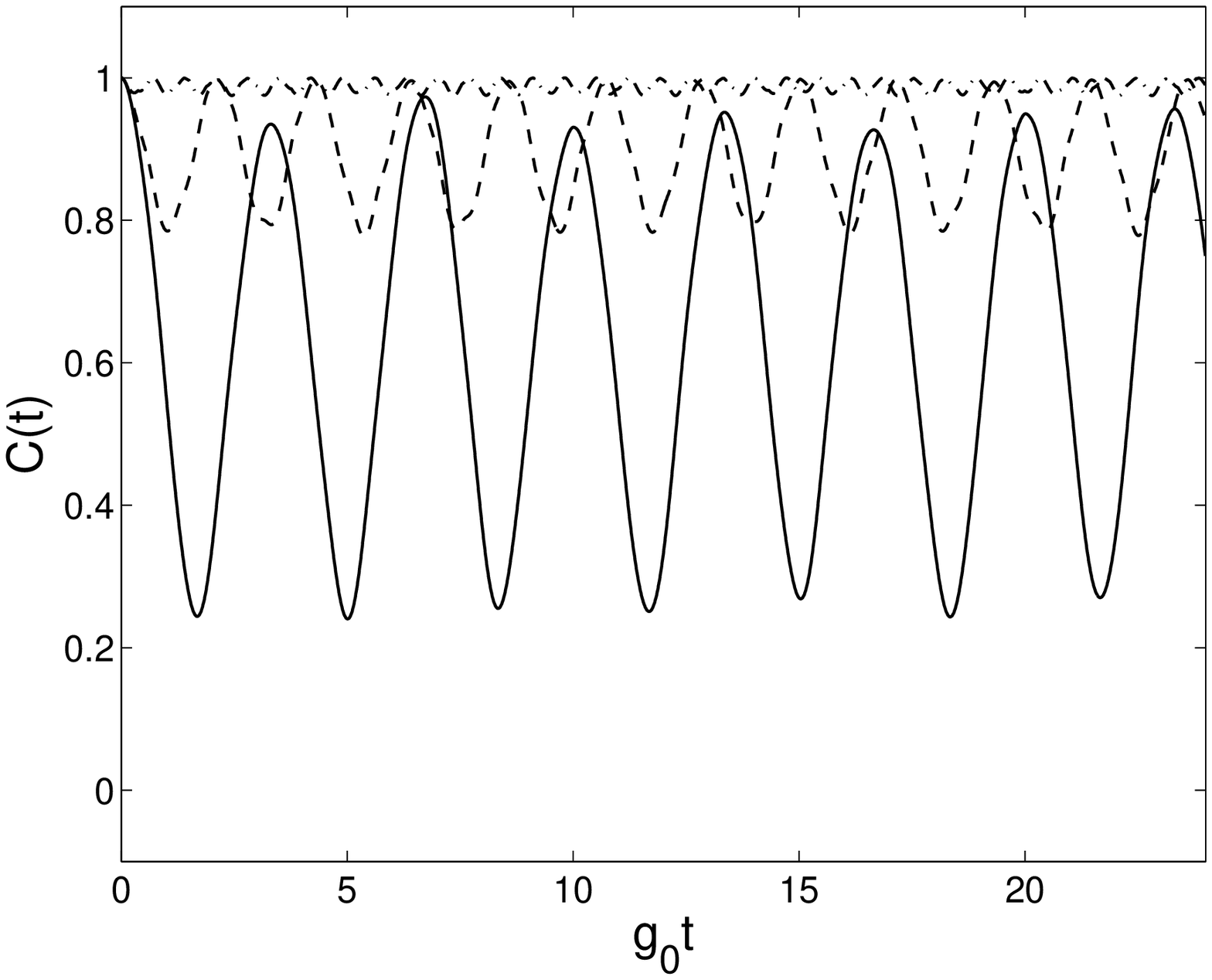}}
\subfigure[$Fd1(t)$]{\label{bellg:Fd0011}
\includegraphics[width=3in]{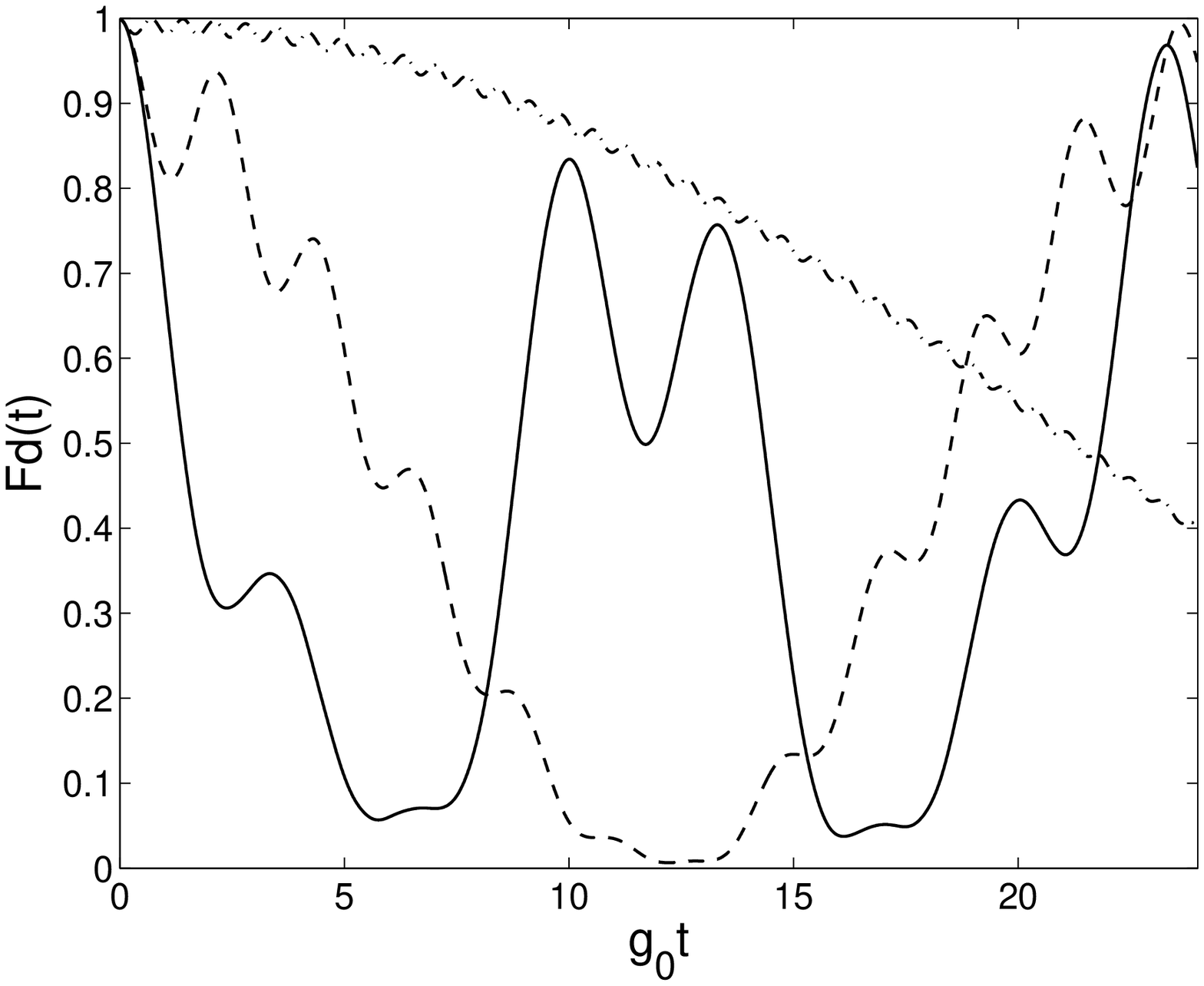}}
\subfigure[$Fd2(t)$]{\label{bellg:Fd1001}
\includegraphics[width=3in]{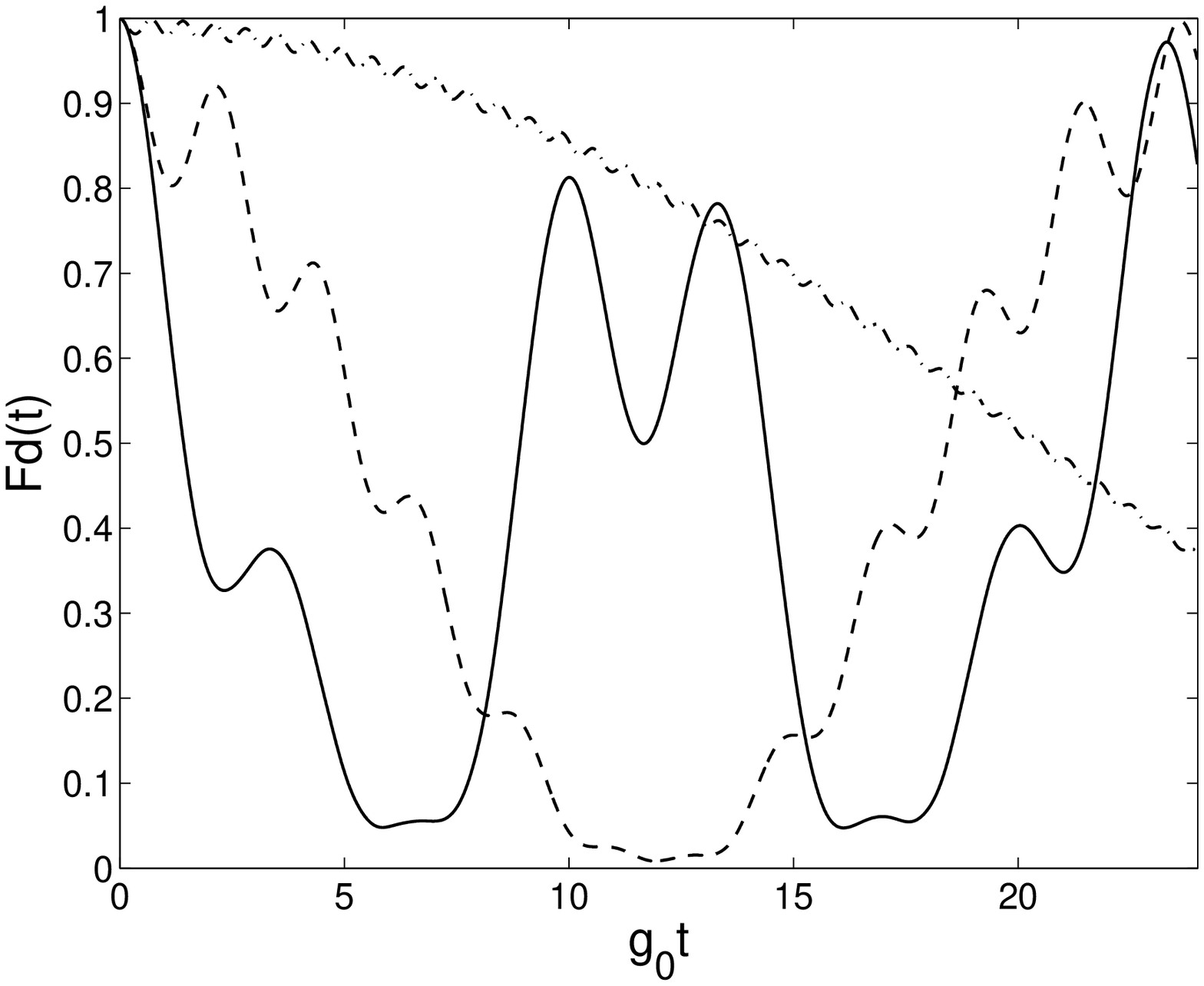}}
\caption{Time evolution for (a) Concurrence from all $4$ Bell
states, (b) Fidelity from $|e_{1,3}\rangle$, (c) Fidelity from
$|e_{2,4}\rangle$ at different values of coupling strength between
subsystem and bath: $g=2g_0$ (solid curve), $g=4g_0$ (dashed
curve), $g=8g_0$ (dot dashed curve). Other parameters are
$N\rightarrow\infty$, $\mu_0=2g_0$, $\gamma=0.6$, $T=g_0$.}
\label{bellg}
\end{figure}

The free evolution (It means the subsystem is decoupled from the
bath or $g_0=0$.) of both concurrence and fidelity of the
subsystem is time-independent, because (i) $H_s$ in Eq. \ref{HS}
($H_s=\mu_0(\sigma_{A}^z+\sigma_{B}^z)$) can not change the
entanglement degree between the two qubits; (ii) when $g_0=0$, the
effect of the bath is excluded from the evolution of the
subsystem. Thus if $g_0=0$, we always have $C(t)=Fd(t)=1$. In Fig.
\ref{bellga:C}, we show different effects on the dynamics by four
anisotropic parameters: $\gamma=0, 0.2, 0.6, 1$. From the four
curves, we obtain two findings. (i) When $\gamma$ is not too
large, the entanglement ($C(0)=1$) can always be recovered to a
high degree after some oscillations. For instances,
$C(\gamma=0,g_0t=8.960)=0.974631$,
$C(\gamma=0.2,g_0t=8.912)=0.961513$,
$C(\gamma=0.6,g_0t=12.672)=0.844158$. Yet the concurrence will
never reach $1$ in a long time scale. (ii) The curve of $\gamma=1$
shows a totally different behavior from the other three cases. It
keeps decreasing with some little fluctuations. Then we turn to
the other sub-figures. Fig. \ref{bellga:Fd1001} is almost the same
as Fig. \ref{bellga:Fd0011}. There are some obvious disagreements
between Fig. \ref{bellga:Fd0011} and Fig. \ref{bellga:C}. For
example, $C(\gamma=0,g_0t=8.960)=0.974631$, but at the same time,
$Fd1(\gamma=0,g_0t=8.960)=0.101803$. In other word, when the
concurrence of the subsystem has been mostly retrieved, the state
of that is not simultaneously back to the its initial state. Only
the combination of the concurrence and the fidelity can give a
complete description of the real revival of the state. It can be
testified in the case of $\gamma=0$ (solid curves in the two
sub-figures). When $g_0t=15.624$, the concurrence evolves to
$C=0.875742$ and the fidelity goes back to $Fd1=0.924277$. So at
that moment, $\rho_s(g_0t)$ is mainly composed by
$|e_{1,3}\rangle\langle{}e_{1,3}|$. \\

In Fig. \ref{bellT:C}, Fig. \ref{bellT:Fd0011} and Fig.
\ref{bellT:Fd1001}, we plot the dynamics of the concurrence and
fidelity at different temperatures. When the temperature is not
too high, such as $T=0.2g_0$ and $T=1g_0$, both concurrence and
fidelity represent a periodical oscillation. At some moments, they
can restore to a high degree. The restoring degree of the both
quantities, however, decreases as the temperature increases.
Similar to Fig. \ref{bellga}, the revivals of the concurrence and
fidelity do not take place simultaneously. When the bath is at a
high temperature, such as $T=5g_0$, the concurrence quickly
declines to zero (to see the dashed line in Fig. \ref{bellT:C})
and does not go back to $C>0$ immediately. It means that when the
local spin bath is adjusted to a high temperature, it makes a
sudden disappearance to the entanglement of a non-localized state
and it will lose the control ability to the subsystem. This is the
effect that has been called ``entanglement sudden death'' (ESD)
\cite{Yu1, Yu2}. In Ref. \cite{Yu2}, after the concurrence goes
abruptly to zero, it arises more or less from nowhere, since there
is no local effect under the action of weak noises. Our model is
still an example of ESD, however, the concurrence arises after
some time due to the local thermal bath.\\

To find out the role of the subsystem-bath coupling strength $g$,
we keep the bath at a moderate temperature $T=g_0$. In Fig.
\ref{bellg:C}, all the three curves show periodical behaviors and
the oscillation amplitudes are strikingly damped by increasing $g$
from $2g_0$ to $8g_0$. When $g$ is up to $8g_0$, the fluctuation
magnitude of concurrence near $C=1$ is too small to be noticed. It
is like the case of $g_0=0$, in which the bath is decoupled from
the subsystem and $C(t)=1$. It is consistent with the claims in
Refs. \cite{TWmodel, Milburn, Jing2} that enough strong
intra-coupling strength among bath spins can make the evolution of
the subsystem be completely determined by the Hamiltonian of
itself $H_s$. But the subsystem state does not receive the same
protection as the subsystem entanglement degree, especially when
$g=8g_0$. Fig. \ref{bellg:Fd0011} manifests that the revival
period of fidelity is much longer than that of concurrence. In
fact, because Qubit $B$ is not under the influence of the bath,
the bath can not make a decoherence-suppression effect on the
subsystem.

\subsection{finite bath spins ($N=40$)}\label{finite}

\begin{figure}[htbp]
\centering \subfigure[$C(t)$]{\label{N40ga:C}
\includegraphics[width=3in]{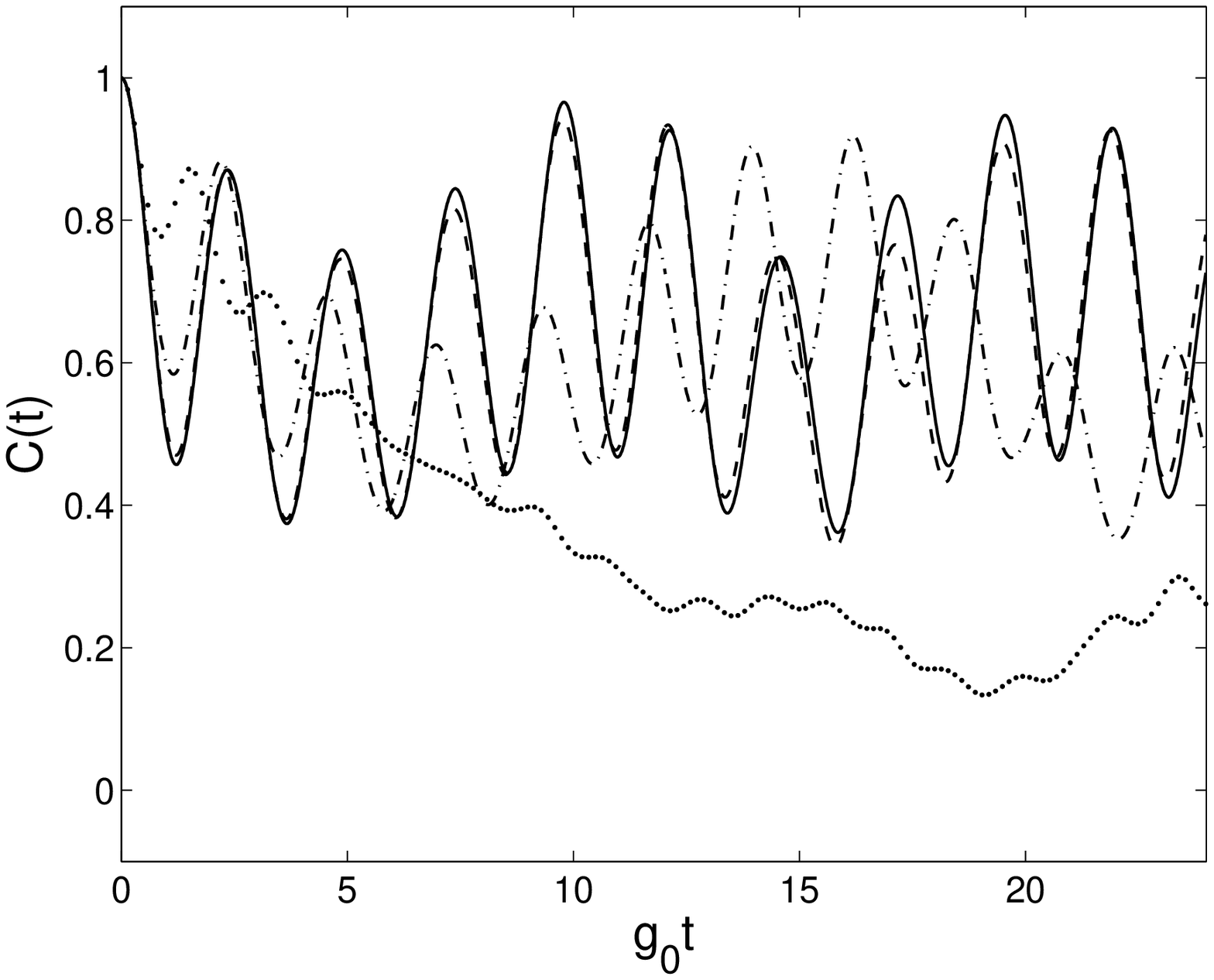}}
\subfigure[$Fd1(t)$]{\label{N40ga:Fd0011}
\includegraphics[width=3in]{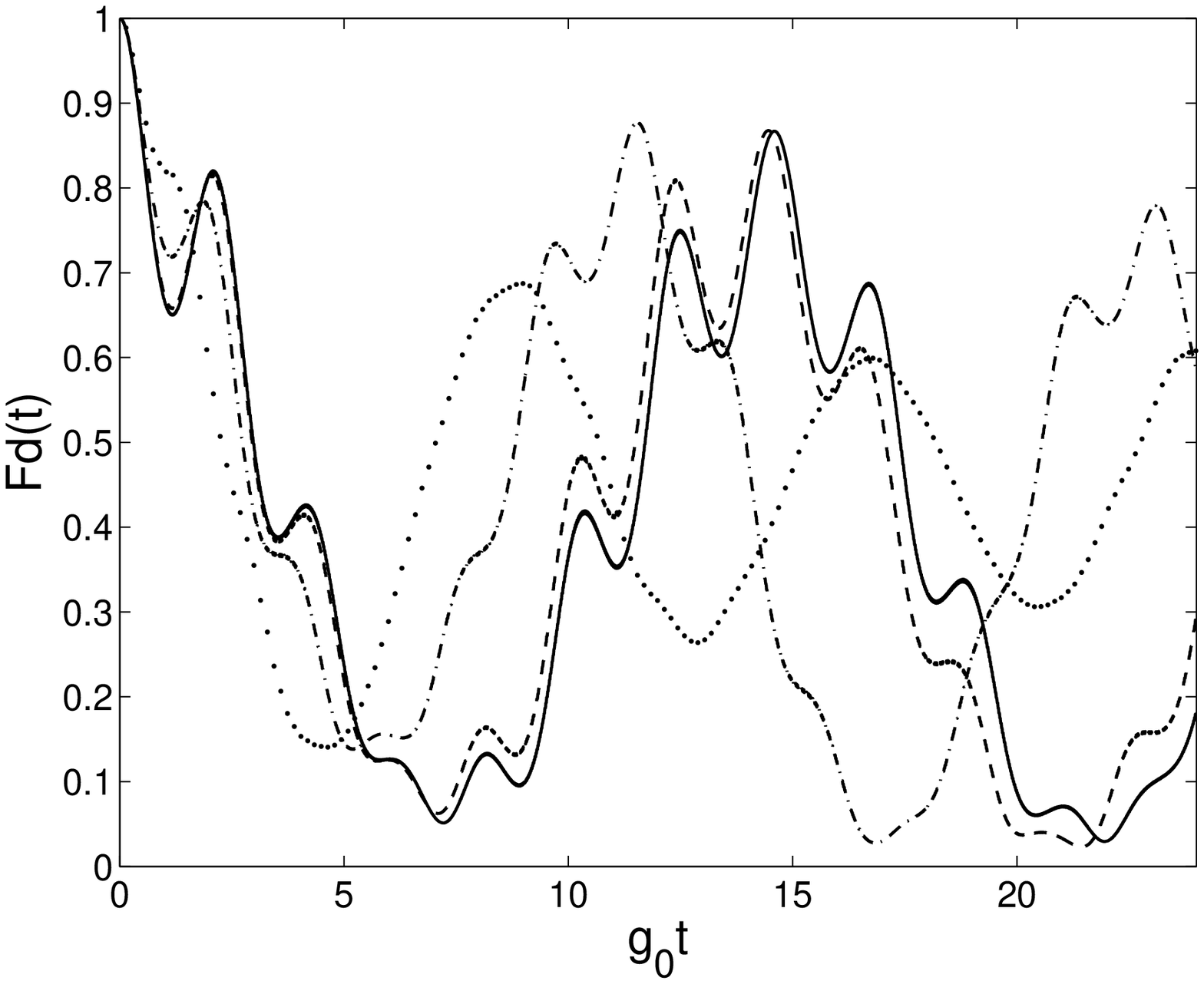}}
\subfigure[$Fd2(t)$]{\label{N40ga:Fd1001}
\includegraphics[width=3in]{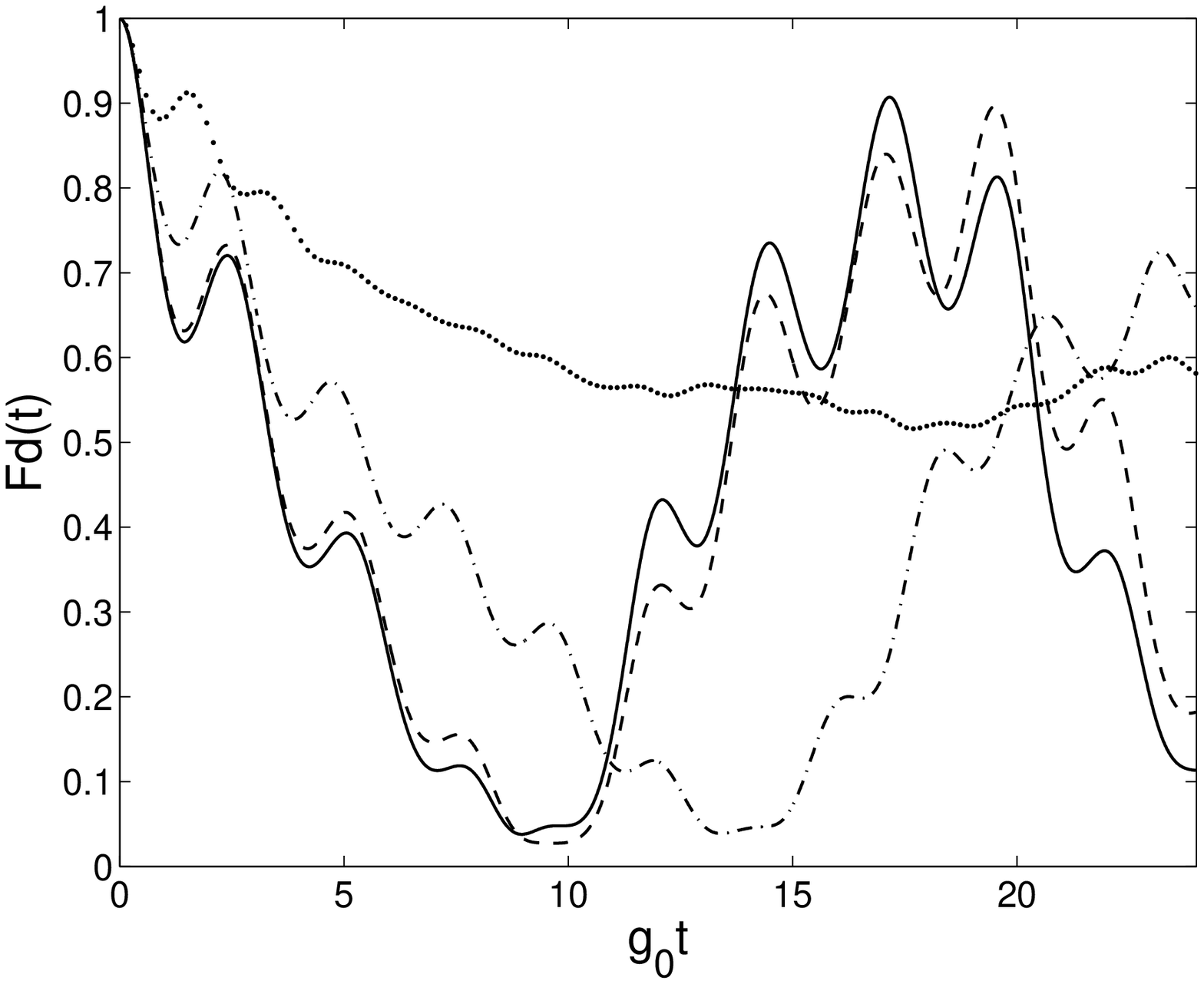}}
\caption{Time evolution for ((a) Concurrence from all $4$ Bell
states, (b) Fidelity from $|e_{1,3}\rangle$, (c) Fidelity from
$|e_{2,4}\rangle$ at different values of anisotropic parameter:
$\gamma=0$ (solid curve), $\gamma=0.2$ (dashed curve),
$\gamma=0.6$ (dot dashed curve), $\gamma=1.0$ (dotted curve).
Other parameters are $N=40$, $\mu_0=2g_0$, $g=g_0$, $T=g_0$.}
\label{N40ga}
\end{figure}

\begin{figure}[htbp]
\centering \subfigure[$C(t)$]{\label{N40T:C}
\includegraphics[width=3in]{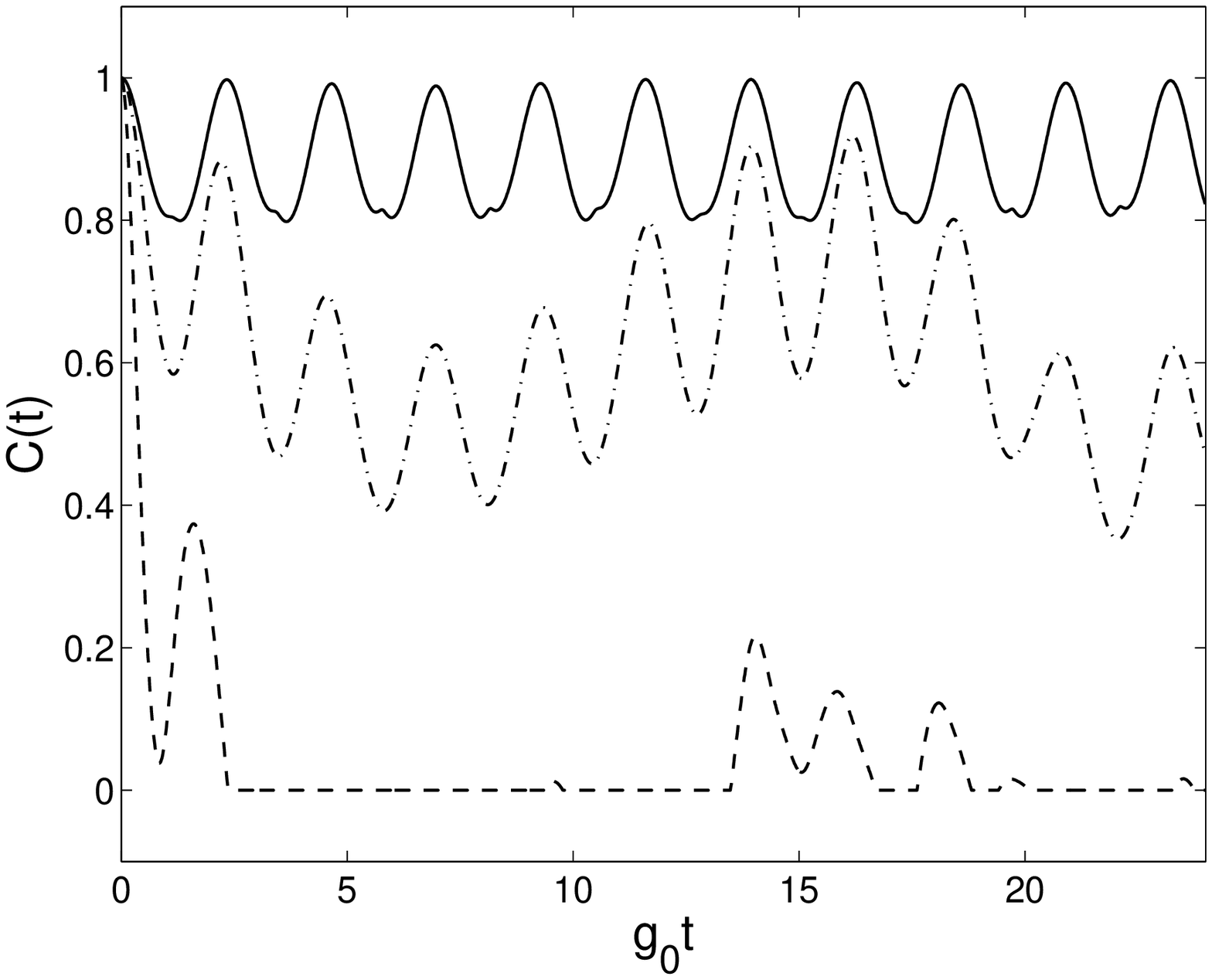}}
\subfigure[$Fd1(t)$]{\label{N40T:Fd0011}
\includegraphics[width=3in]{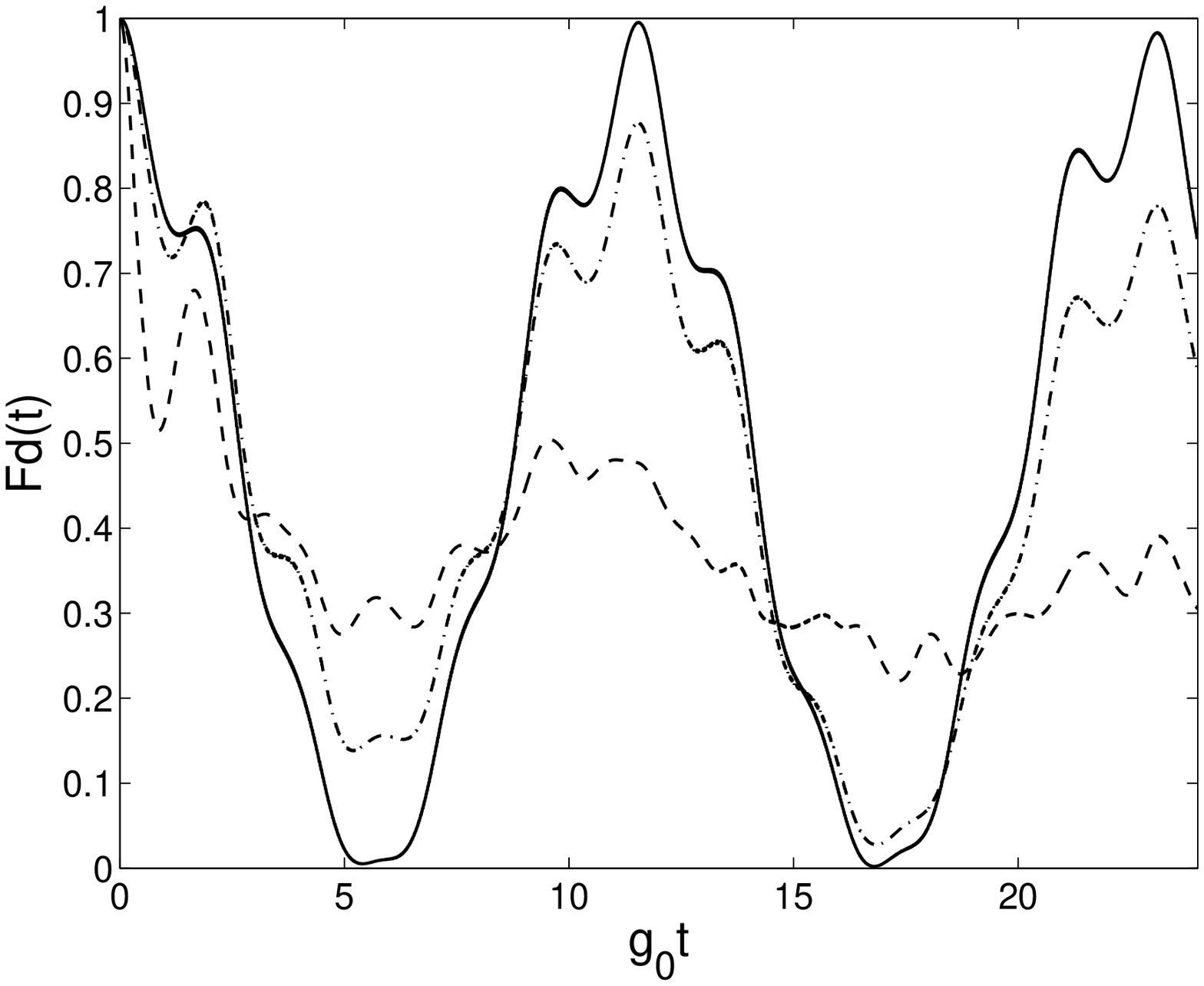}}
\subfigure[$Fd2(t)$]{\label{N40T:Fd1001}
\includegraphics[width=3in]{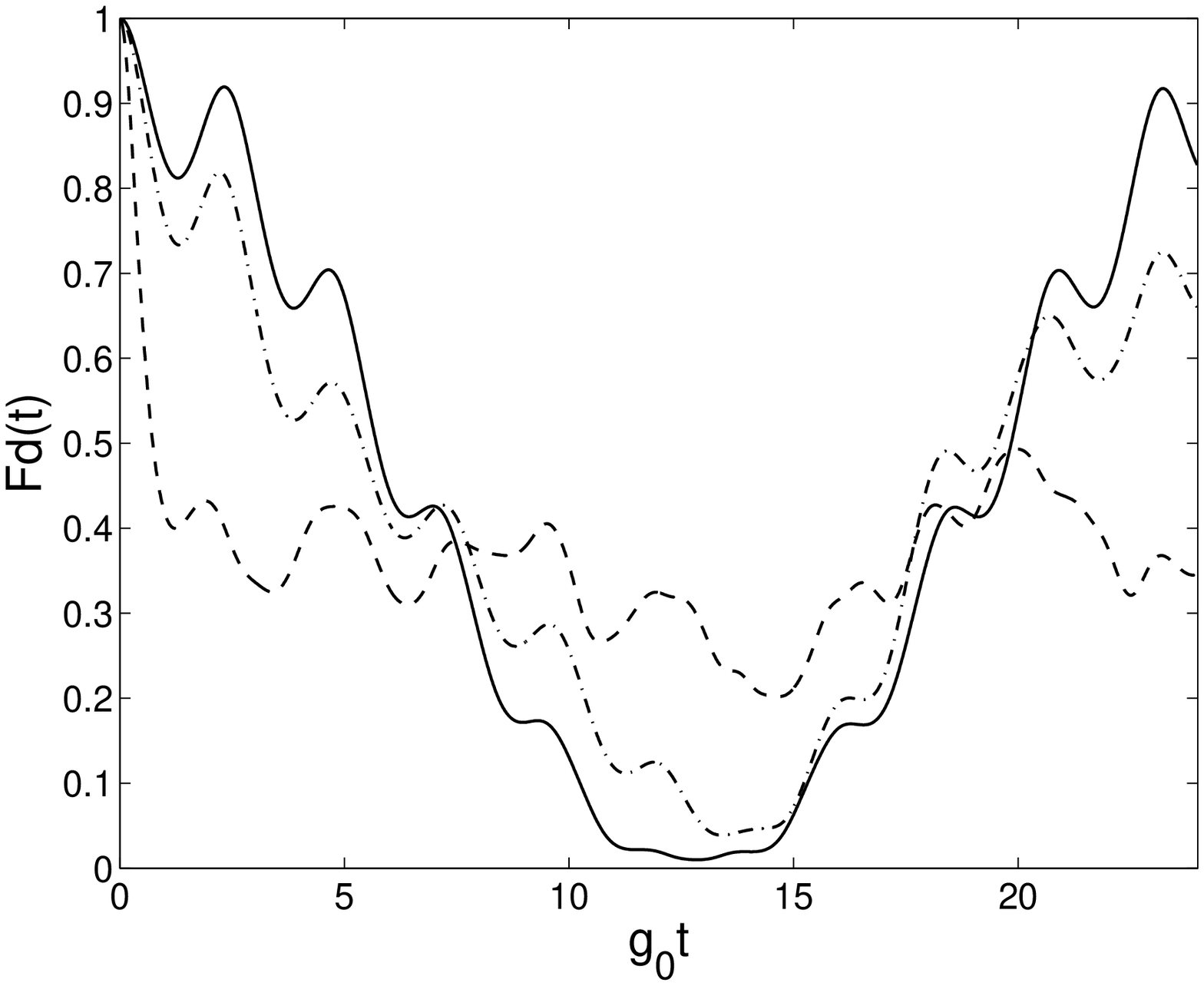}}
\caption{Time evolution for (a) Concurrence from all $4$ Bell
states, (b) Fidelity from $|e_{1,3}\rangle$, (c) Fidelity from
$|e_{2,4}\rangle$ at different values of temperature: $T=0.2g_0$
(solid curve), $T=g_0$ (dot dashed curve), $T=5g_0$ (dashed
curve). Other parameters are $N=40$, $\mu_0=2g_0$, $g=g_0$,
$\gamma=0.6$.} \label{N40T}
\end{figure}

\begin{figure}[htbp]
\centering \subfigure[$C(t)$]{\label{N40g:C}
\includegraphics[width=3in]{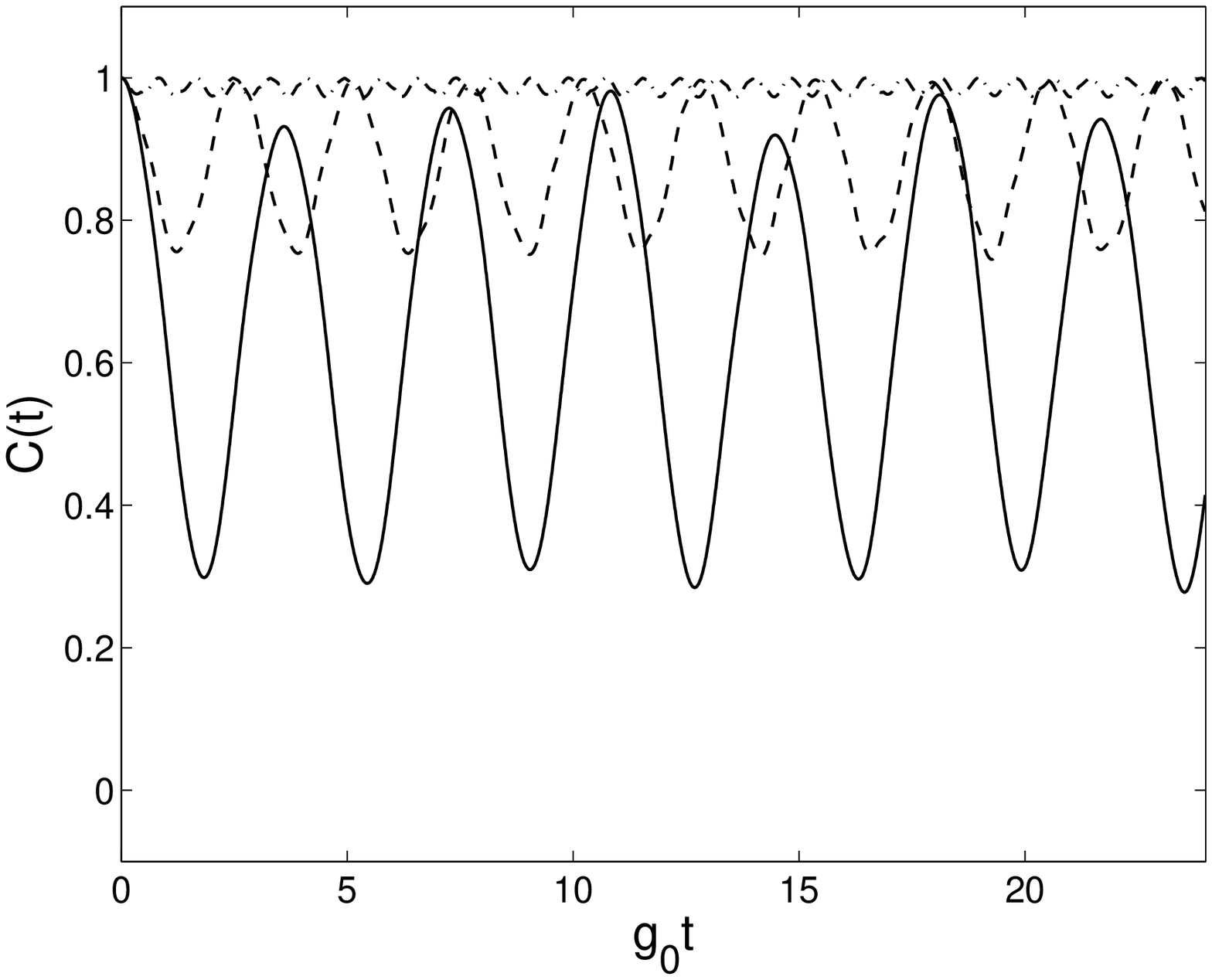}}
\subfigure[$Fd1(t)$]{\label{N40g:Fd0011}
\includegraphics[width=3in]{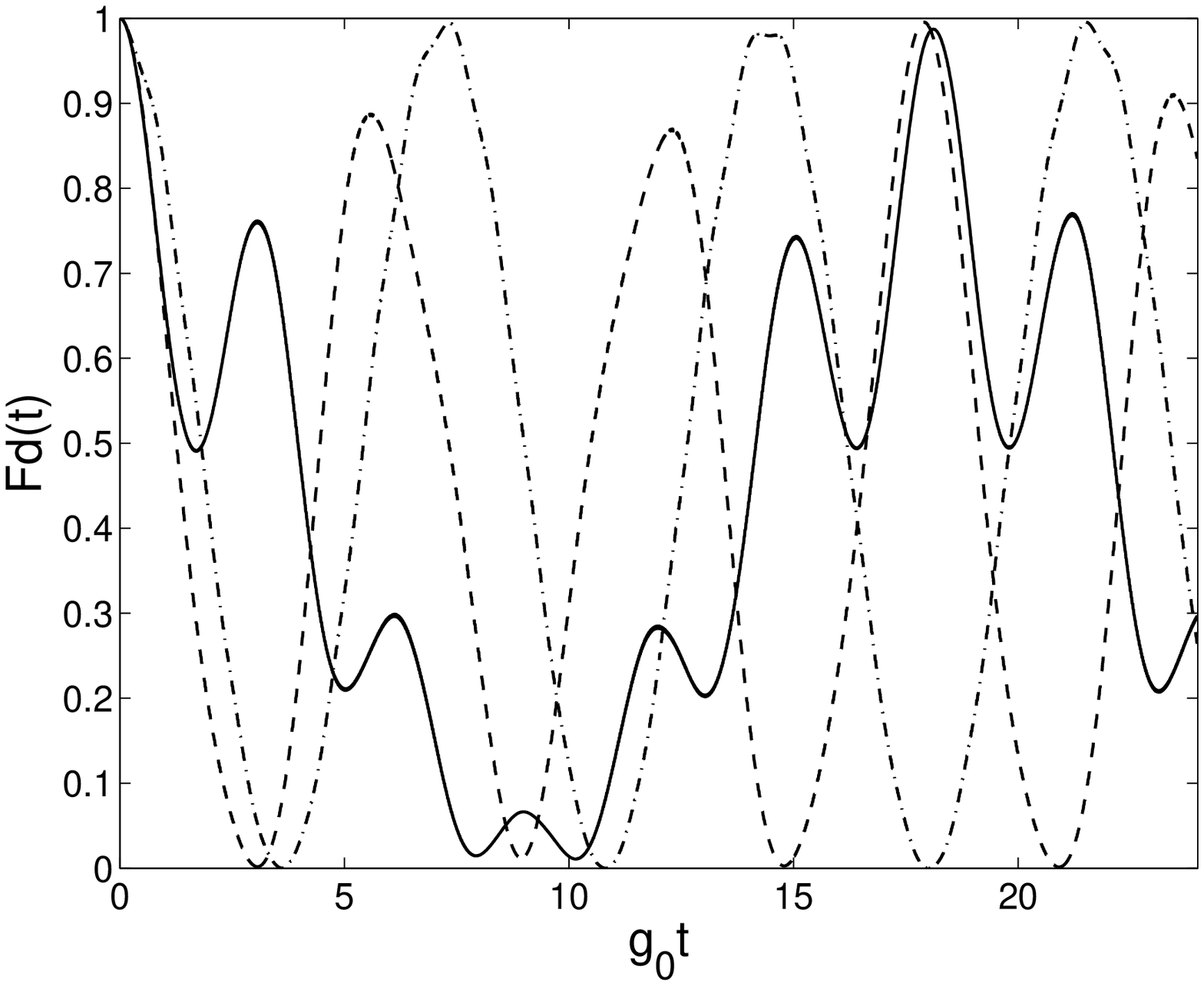}}
\subfigure[$Fd2(t)$]{\label{N40g:Fd1001}
\includegraphics[width=3in]{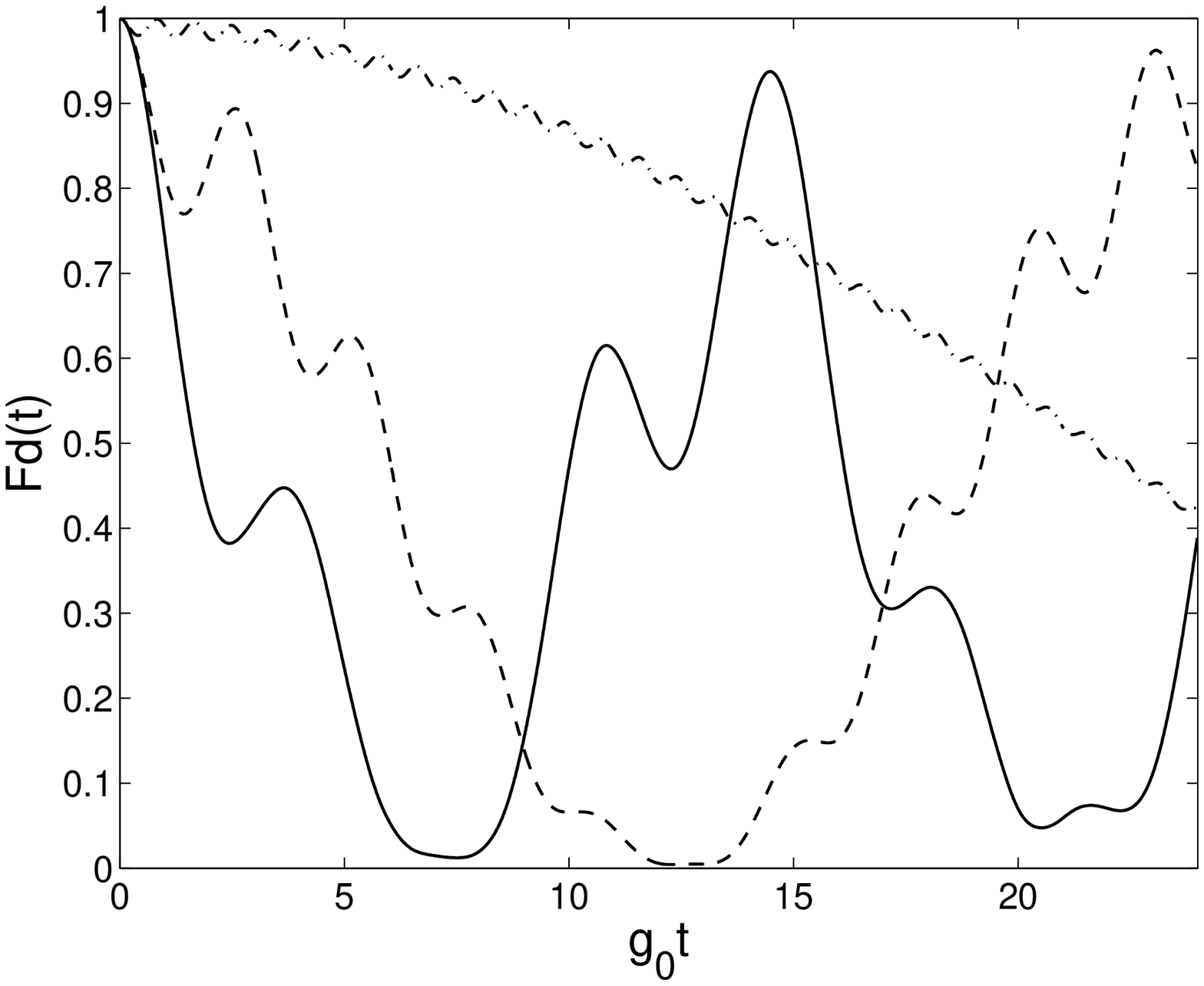}}
\caption{Time evolution for (a) Concurrence from all $4$ Bell
states, (b) Fidelity from $|e_{1,3}\rangle$, (c) Fidelity from
$|e_{2,4}\rangle$ at different values of coupling strength between
subsystem and bath: $g=2g_0$ (solid curve), $g=4g_0$ (dashed
curve), $g=8g_0$ (dot dashed curve). Other parameters are $N=40$,
$\mu_0=2g_0$, $\gamma=0.6$, $T=g_0$.} \label{N40g}
\end{figure}

In the previous two-center-spin-spin-bath works \cite{Yuan,
Jing3}, it is supposed that the number of bath spins is infinite,
which helps to reduce the Hamiltonian Eqs. \ref{H_SB} and
\ref{H_B} to a simple form. Yet as the controlling device in a
real quantum information equipment, the spin bath, in principle,
should be made of finite number of spins-1/2. Then in this
subsection, we use the $1$st order expansion of the Hamiltonian
(Eqs. \ref{H_SB} and \ref{H_B}) to introduce a finite $N$ in the
present problem. The error about this approximation is about
$O(1/N^2)$ as Eq. \ref{H_B} indicates. Without loss of generality,
we set $N=40$ . \\

Comparing the result of Fig. \ref{bellga} with that of Fig.
\ref{N40ga}, we can find some agreements and some disagreements.
The most identical characteristic between them is that the
concurrence dynamics is independent of the choice of state as long
as it is one of the four Bell states. Yet when $\gamma=1$, the
concurrence (dotted curve) does not decrease monotonously during
the given time. For fidelity, it is shown that with a bigger
anisotropic parameter $\gamma$, the curves in Fig.
\ref{N40ga:Fd0011} oscillate with a shorter period, which is
opposite to the tendency in Fig. \ref{bellga:Fd0011}. While Fig.
\ref{N40ga:Fd1001} is almost the same as Fig. \ref{bellga:Fd1001},
with a little longer oscillation period. It seems that the
fidelity evolution of the $2$nd group is not very sensitive to the
bath-spin number $N$. The differences between the infinite $N$ and
finite $N$ cases might arise from their different energy-level
numbers and corresponding weights in our numerical scheme (in
subsection \ref{finite}). Under the same requirement of numerical
accuracy, for the infinite $N$, we need to consider $14, 15, 18$
and $20$ energy levels when $\gamma$ is $0, 0.2, 0.6$ and $1$
respectively; for $N=40$, we calculate $9, 10, 17$ and $18$ levels
respectively. \\

In the comparison of Fig. \ref{N40T} with Fig. \ref{bellT}, we can
also find the effect of a finite $N$. At low temperature
($T=0.2g_0$), the entanglement degree of the subsystem qubits
oscillates with a nearly perfect period between the value of $0.8$
and $1.0$ (the solid curve in Fig. \ref{N40T:C}). However, the
subsystem of Group $1$ (Group $2$) goes back to its own initial
state only once in almost five (ten) revival periods of
concurrence, which is illustrated by the corresponding curve in
Fig. \ref{N40T:Fd0011} (Fig. \ref{N40T:Fd1001}). It is obvious
that the increase of the temperature will also destroy this
perfect oscillation. When $T=1g_0$, the second peak value,
$C(g_0t=16.19)=0.918798$ is lower than the first one $C(g_0t=0)=1$
(to see the dot dashed curve in Fig. \ref{N40T:C}) and the peak
$Fd1(g_0t=11.55)=0.877301$ is higher than
$Fd1(g_0t=23.13)=0.779953$ (to see the dot dashed curve in
\ref{N40T:Fd0011}). When the temperature is up to $5g_0$, the
entanglement vanishes to zero in a fairly short stretch of time.
The entanglement ``death'' time is longer than that in the case of
$N\rightarrow\infty$ (Fig. \ref{bellT}). So whether
$N\rightarrow\infty$ or $N$ is finite, the revival of the
concurrence after ESD results from the effect of thermal bath. \\

For the subsystem-bath coupling $g$, the dynamics of concurrence
in the case of $N=40$ (to see Fig. \ref{N40g:C}) is almost the
same as that in the $N\rightarrow\infty$ case (to compare it with
Fig. \ref{bellg:C}). The evolution of the fidelity of the $1$st
group shows significant changes when $N$ is changed from infinity
(Fig. \ref{bellg:Fd0011}) to $40$ (Fig. \ref{N40g:Fd0011}) while
the fidelity of the $2$nd group does not show very obvious
changes. And for the $1$st group, all the three cases behave
periodical oscillations. They manifest that the subsystem in the
condition of $N=40$ can be restored to the initial state with more
chances or possibilities than that in the condition of
$N\rightarrow\infty$.

\section{Conclusion}\label{conclusion}

We studied the time evolution of two separated qubit spins with a
thermal equilibrium bath composed of infinite or finite spins in a
quantum anisotropic Heisenberg XY model. The bath can be treated
effectively as a single pseudo-spin of $N/2$ according to the
symmetry of the Hamiltonian. By the Holstein-Primakoff
transformation and the first order of $1/N$ expansion, it is
further considered as a single-mode boson field. The pair of
qubits served as an quantum information device is initially
prepared in a Bell state. It is interesting that the concurrence
and the fidelity dynamics of the subsystem can be controlled by
some characteristic parameters of the spin bath. Through the
adjustment, we show that (i) the concurrence dynamics of the
subsystem is independent of the initial state, whether $N$ is
infinite or finite, however, the fidelity dynamics is divided into
two groups; (ii) smaller anisotropic parameter $\gamma$ can help
the subsystem to evolve into a highly-entangled state, but this
restoration should be measured by the combination of concurrence
and fidelity; (iii) the bath at higher temperature makes a sudden
death to the entanglement (ESD) of the subsystem and strongly
destroys the fidelity of that; (iv) the spin-bath can help to keep
the high entanglement degree between the two subsystem spins in
the condition of large intra-coupling $g$.

\begin{acknowledgments}

We would like to acknowledge the support from the National Natural
Science Foundation of China under grant No. 10575068, the Natural
Science Foundation of Shanghai Municipal Science Technology
Commission under grant Nos. 04ZR14059 and 04dz05905 and the CAS
Knowledge Innovation Project Nos. KJcx.syw.N2.

\end{acknowledgments}

\end{document}